\documentclass[prl,twocolumn,showkeys,showpacs,superscriptaddress]{revtex4}
\usepackage{epsfig,latexsym,amssymb}

\begin{document}

\title{Critical Scaling of Shear Viscosity at the Jamming Transition}

\author{Peter Olsson}
\affiliation{Department of Physics, Ume{\aa} University, 901 87 Ume{\aa}, Sweden}
\author{S. Teitel}
\affiliation{Department of Physics and Astronomy, University of
Rochester, Rochester, NY 14627}
\date{\today}

\begin{abstract}
We carry out numerical simulations to study transport behavior about the jamming transition of a model granular material in two dimensions at zero temperature.  Shear viscosity $\eta$ is computed as a function of particle volume density $\rho$ and applied shear stress $\sigma$, for diffusively moving particles with a soft core interaction.  We find an excellent scaling collapse of our data as a function of the scaling variable $\sigma/|\rho_{\rm c}-\rho|^\Delta$, where $\rho_{\rm c}$ is the critical density at $\sigma=0$ (``point {\it J}"), and $\Delta$ is the crossover scaling critical exponent.  We define a correlation length $\xi$ from velocity correlations in the driven steady state, and show that it diverges at point {\it J}.  Our results support the assertion that jamming is a true second order critical phenomenon. 
\end{abstract}
\maketitle
 
In granular materials, or other spatially disordered systems such as colloidal glasses, gels, and foams, in which thermal fluctuations are believed to be negligible, a {\it jamming transition} has been proposed: upon increasing the volume density (or ``packing fraction") of particles $\rho$ above a critical $\rho_{\rm c}$, the sudden appearance of a finite shear stiffness signals a transition between flowing liquid and rigid (but disordered) solid states \cite{LiuNagel1}.  It has further been proposed by Liu and Nagel and co-workers \cite{LiuNagel2,OHern2} that this jamming transition is a special second order critical point (``point {\it J}'') in a wider phase diagram whose axes are volume density $\rho$, temperature $T$, and applied shear stress $\sigma$ (the latter parameter taking one out of equilibrium to non-equilibrium driven steady states).  
A surface in this three dimensional parameter space then separates jammed from flowing states,
and the intersection of this surface with the equilibrium $\rho-T$ plane at $\sigma=0$ is related to the structural glass transition.

Several numerical \cite{OHern2,Durian,Makse,OHern1,Drocco,Silbert,Ellenbroek,Ning}, theoretical \cite {Schwarz,Fisher,Chak,Wyart} and experimental \cite{Makse,Weitz,Majmudar,Durian2,Swinney} works have investigated the jamming transition, 
mostly by considering behavior as the transition is approached from the {\it jammed} side.
In this work we consider the {\it flowing} state, computing the shear viscosity $\eta$ under applied uniform shear stress.  
Previous works have simulated the flowing response to applied shear in glassy systems at finite temperature \cite{Yamamoto,Berthier, Varnik}, and in foams \cite{Durian} and granular systems \cite{Ning} at $T=0$, $\rho>\rho_c$.
Here we consider the $\rho-\sigma$ plane at $T=0$, showing for the first time that, near point {\it J}, $\eta^{-1}(\rho,\sigma)$ collapses to a universal scaling function of the variable $\sigma/|\rho_{\rm c}-\rho|^\Delta$ for both $\rho<\rho_{\rm c}$ and $\rho>\rho_{\rm c}$.   
We further define a correlation length $\xi$ from steady state velocity correlations, and show that it diverges at point {\it J}.
Our results support that jamming is a true second order critical phenomenon.

Following  O'Hern et al. \cite{OHern2}, we simulate frictionless soft disks in two dimensions (2D) using a bidisperse mixture with equal numbers of disks of two different radii. The  radii ratio is 1.4 and the interaction between the particles is,
\begin{equation}
V(r_{ij})=\left\{ 
\begin{array}{ll}\epsilon(1-r_{ij}/d_{ij})^2/2 &{\rm for}\quad r_{ij}<d_{ij}\\ 0 &{\rm for}\quad r_{ij}\ge d_{ij}\end{array}\right.
\end{equation}
where $r_{ij}$ is the distance between the centers of two particles $i$ and $j$, and $d_{ij}$ is the sum of their radii. Particles are non-interacting when they do not touch, and interact with a harmonic repulsion when they overlap.  We measure length in units such that the smaller diameter is unity, and energy in units such that $\epsilon=1$.  A system of $N$ disks in an area $L_x\times L_y$ thus has a volume density 
\begin{equation}
\rho = N\pi(0.5^2+0.7^2)/(2L_xL_y)\enspace.  
\label{erho}
\end{equation}

To model an applied uniform shear stress, $\sigma$, we first use Lees-Edwards boundary conditions \cite{LeesEdwards} to introduce a uniform shear {\it strain}, $\gamma$.  Defining particle $i$'s position as ${\bf r}_i=(x_i+\gamma y_i,y_i)$, we apply periodic boundary conditions on the coordinates $x_i$ and $y_i$ in an $L_x\times L_y$ system.  In this way, each particle upon mapping back to itself under the periodic boundary condition in the $\hat y$ direction, has displaced a distance $\Delta x=\gamma L_y$ in the $\hat x$ direction, resulting in a shear strain $\Delta x/L_y=\gamma$.  When particles do not touch, and hence all mutual forces vanish,  $x_i$ and $y_i$ are constant and a time dependent strain $\gamma(t)$ produces a uniform shear flow, $d{\bf r}_i/dt = y_i(d\gamma/dt)\hat x$.  
When particles touch, we assume a diffusive response to the inter-particle forces, as would be appropriate if the particles were immersed in a highly viscous liquid or resting upon a rough surface with high friction.  This results in the following equation of motion, which was first proposed as a model for sheared foams \cite{Durian},
\begin{equation}
{d{\bf r}_i\over dt}=-D\sum_j {dV(r_{ij})\over d{\bf r}_i}+y_i{d\gamma\over dt}\hat x\enspace.
\label{edrdt}
\end{equation}
The strain $\gamma$ is then treated as a dynamical variable, obeying the equation of motion,
\begin{equation}
{d\gamma\over dt} = D_\gamma\left[L_xL_y\sigma -\sum_{i\ne j}{dV(r_{ij})\over d\gamma}\right]\enspace,
\label{edgdt}
\end{equation}
where the applied stress $\sigma$ acts like an external force on $\gamma$ and the interaction terms $V(r_{ij})$ depend on $\gamma$ via the particle separations, ${\bf r}_{ij}=([x_i-x_j]_{L_x}+\gamma [y_i-y_j]_{L_y},[y_i-y_j]_{L_y})$, where by $[\dots ]_{L_\mu}$ we mean that the difference is to be taken, invoking periodic boundary conditions, so that the result lies in the interval $(-L_\mu/2, L_\mu/2]$.  The constants $D$ and $D_\gamma$ are set by the dissipation of the medium in which the particles are embedded; we take units of time such that $D=D_\gamma\equiv 1$.

In a flowing state at finite $\sigma>0$, the sum of the interaction terms is of order $O(N)$ so that the right hand side of Eq.~(\ref{edgdt}) is $O(1)$.  The strain $\gamma(t)$ increases linearly in time on average, leading to a sheared flow of the particles with average velocity gradient  
$dv_x/dy = \langle d\gamma/dt\rangle$, where $v_x(y)$ is the average velocity in the $\hat x$ direction of the particles at height $y$.  We then measure the shear viscosity, defined by,
\begin{equation}
\eta \equiv {\sigma\over dv_x/dy}={\sigma\over \langle d\gamma/dt\rangle}\enspace.
\end{equation}
We expect $\eta^{-1}$ to vanish in a jammed state.

We integrate the equations of motion, Eqs.~(\ref{edrdt})-(\ref{edgdt}), starting from an initial random configuration, using the Heuns method.  The time step $\Delta t$ is varied according to system size to ensure our results are independent of $\Delta t$.  We consider a fixed number of particles $N$, in a square system $L\equiv L_x=L_y$, and vary the volume density $\rho$ by adjusting the length $L$ according to Eq.~(\ref{erho}).  We simulate for times $t_{\rm tot}$ such that the total relative displacement per unit length transverse to the
direction of motion is typically $\gamma(t_{\rm tot})\sim 10$, with $\gamma(t_{\rm tot})$ ranging between $1$ and $200$ depending on the particular system parameters.

\begin{figure}[tbp]
\epsfxsize=8.6truecm
\epsfbox{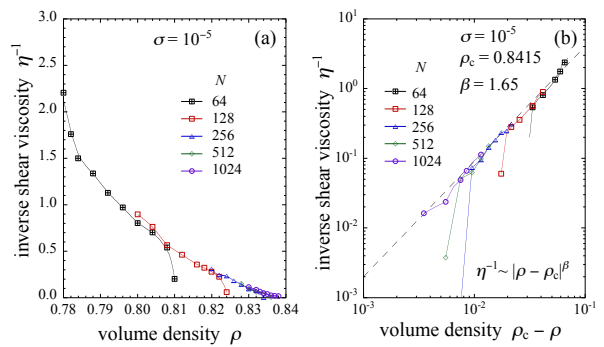}
\caption{(color online) 
a) Plot of inverse shear viscosity $\eta^{-1}$ vs volume density $\rho$ for several different numbers of particles $N$, at constant small applied shear stress $\sigma=10^{-5}$.  As $N$ increases, one see jamming at a limiting value of the density $\rho_{\rm c}\sim 0.84$.
b) Log-log replot of the data of (a) as $\eta^{-1}$ vs $\rho_{\rm c}-\rho$, with $\rho_{\rm c}=0.8415$.  The dashed line has slope $\beta=1.65$ indicating the continuous algebraic vanishing of $\eta^{-1}$ at $\rho_{\rm c}$ with a critical exponent $\beta$.
}
\label{f1}
\end{figure}

In Fig.~\ref{f1} we show our results for $\eta^{-1}$ using a fixed small shear stress, $\sigma=10^{-5}$, representative of the $\sigma\to 0$ limit.  Our raw results are shown in Fig.~\ref{f1}a for several different numbers of particles $N$ from $64$ to $1024$.
Comparing the curves for different $N$ as $\rho$ increases, we see that they overlap for some range of $\rho$, before each drops discontinuously into a jammed state.
As  $N$ increases, the onset value of $\rho$ for jamming increases to a limiting value $\rho_{\rm c}\simeq 0.84$ (consistent with the value for random close packing \cite{OHern2}) and $\eta^{-1}$ vanishes continuously.  For finite $N$, systems jam below $\rho_{\rm c}$ because there is always a finite probability to find a configuration with a force chain spanning the width of the system, thus causing it to jam; and at $T=0$, once a system jams, it remains jammed for all further time.  As the system evolves dynamically with increasing simulation time, it explores an increasing region of configuration space, and ultimately finds a configuration that causes it to jam.  The statistical weight of such jamming configurations decreases, and hence the average time required to jam increases, as one either decreases $\rho$, or increases $N$ \cite{OHern2}.  In the limit $N\to\infty$, we expect jamming will occur in finite time only for $\rho\ge\rho_{\rm c}$.  In Fig.~\ref{f1}b we show a log-log plot of $\eta^{-1}$ vs $\rho_{\rm c}-\rho$, using a value $\rho_{\rm c}=0.8415$.  We see that the data in the unjammed state is well approximated by a straight line of slope $\beta=1.65$, giving
$\eta^{-1}\sim |\rho -\rho_{\rm c}|^\beta$ in agreement with the expectation that point {\it J} is a second order phase transition.

If point {\it J} is indeed a true critical point, one expects that its influence will be felt also at finite values of the stress $\sigma$, with $\eta^{-1}$ obeying a typical scaling law,
\begin{equation}
\eta^{-1}(\rho,\sigma)=|\rho-\rho_{\rm c}|^\beta f_{\pm}\left({\sigma\over|\rho-\rho_{\rm c}|^\Delta}\right)\enspace.
\label{escale}
\end{equation}
Here $z\equiv \sigma/|\rho-\rho_{\rm c}|^\Delta$ is the crossover scaling variable, $\Delta$ is the crossover scaling critical exponent, and $f_-(z)$, $f_+(z)$ are the two branches of the crossover scaling function for $\rho<\rho_{\rm c}$ and $\rho>\rho_{\rm c}$ respectively.  

\begin{figure}[tbp]
\epsfxsize=8.6truecm
\epsfbox{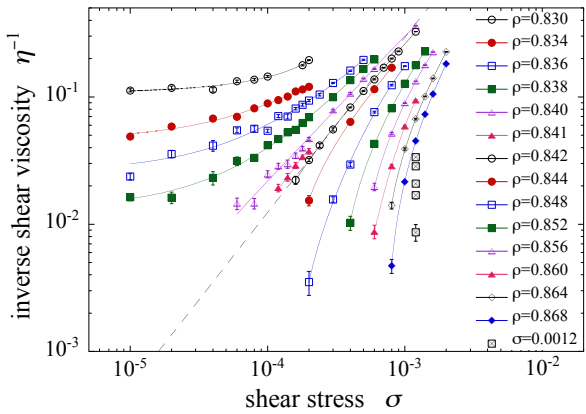}
\caption{(color online) 
Plot of inverse shear viscosity $\eta^{-1}$ vs applied shear stress $\sigma$ for several different values of the volume density $\rho$.  The dashed line represents the power law dependence expected exactly at $\rho=\rho_{\rm c}$ and has a slope $\beta/\Delta = 1.375$. Solid lines are guides to the eye.  Points labeled $\sigma=0.0012$ correspond to densities $\rho=0.870$, $0.872$, $0.874$, $0.876$, and $0.878$.
}
\label{f2}
\end{figure}

In Fig.~\ref{f2} we show a log-log plot of  inverse shear viscosity $\eta^{-1}$ vs applied shear stress $\sigma$, for several different values of volume density $\rho$.  Our results are for systems large enough that we believe finite size effects are negligible.  We use $N=1024$ for $\rho<0.844$ and $N=2048$ for $\rho\ge 0.844$.  Again we see that $\rho_{\rm c}\simeq 0.8415$ separates two limits of behavior.  For $\rho<\rho_{\rm c}$, $\log \eta^{-1}$ is convex in $\log \sigma$, decreasing to a finite value as $\sigma\to 0$.  For $\rho>\rho_{\rm c}$, $\log \eta^{-1}$ is concave in $\log \sigma$, decreasing towards zero as $\sigma\to 0$.  The dashed straight line,
separating the two regions of behavior, indicates the power law dependence that is expected exactly at $\rho=\rho_{\rm c}$ (see below).  Similar power law behavior at $\rho_c$ was recently found in simulations of a three dimensional granular material \cite{Sasa}.

\begin{figure}[tbp]
\epsfxsize=8.6truecm
\epsfbox{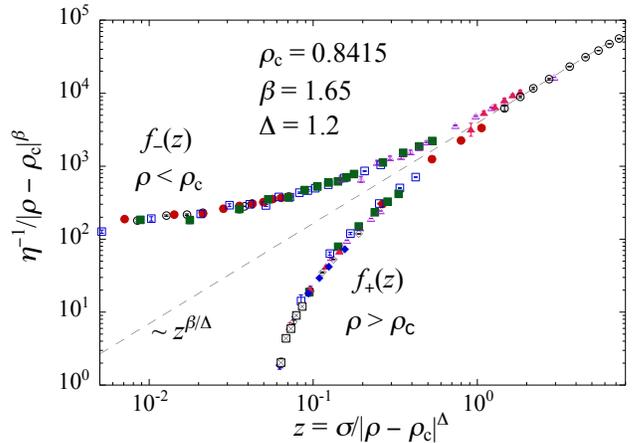}
\caption{(color online) 
Plot of scaled inverse viscosity $\eta^{-1}/|\rho-\rho_{\rm c}|^\beta$ vs scaled shear stress $z\equiv\sigma/|\rho-\rho_{\rm c}|^\Delta$ for the data of Fig.~\ref{f2}.  We find an excellent collapse to the scaling form of Eq.~(\ref{escale}) using values $\rho_{\rm c}=0.8415$, $\beta=1.65$ and $\Delta=1.2$.  The dashed line represents the large $z$ asymptotic dependence, $\sim z^{\beta/\Delta}$.  
Data point symbols correspond to those used in Fig.~\ref{f2}.
}
\label{f3}
\end{figure}

In Fig.~\ref{f3} we replot the data of Fig.~\ref{f2} in the scaled variables  $\eta^{-1}/|\rho-\rho_{\rm c}|^\beta$ vs $\sigma/|\rho-\rho_{\rm c}|^\Delta$.  Using $\rho_{\rm c}=0.8415$, $\beta=1.65$ (the same values used in Fig.~\ref{f1}b) and $\Delta=1.2$, we find an excellent scaling collapse in agreement with the prediction of Eq.~(\ref{escale}).
As the scaling variable $z\to 0$, $f_-(z)\to$ constant; this gives the vanishing of $\eta^{-1}\sim |\rho-\rho_{\rm c}|^\beta$ at $\sigma=0$.  As $z\to\infty$, both branches of the scaling function approach a common curve, $f_\pm (z)\sim z^{\beta/\Delta}$, so that precisely at $\rho=\rho_{\rm c}$, $\eta^{-1}\sim\sigma^{\beta/\Delta}$ as $\sigma\to 0$ \cite{nonlinear}.  
This is shown as the dashed line in both Figs.~\ref{f3} and \ref{f2}.
A similar scaling collapse of $\eta$ has been found in simulations \cite{Berthier} of a sheared Lennard-Jones glass, 
as a function of temperature and applied shear strain rate $\dot\gamma$, but only above the glass transition, $T>T_c$.
By comparing the goodness of the scaling collapse as parameters are varied, we estimate the 
accuracy of the critical exponents to be roughly, $\beta=1.7\pm 0.2$ and $\Delta =1.2\pm 0.2$.

That the crossover scaling exponent $\Delta>0$, implies that $\sigma$ is a relevant variable in the renormalization group sense, and that critical behavior at finite $\sigma$ should be in a different universality class from the jamming transition at point {\it J} (i.e. $\sigma=0$).  The nature of jamming at finite $\sigma>0$ will be determined by the behavior of the branch of the crossover scaling function $f_+(z)$, that describes behavior for $\rho>\rho_c$.  From Fig.~\ref{f3} we see that $f_+(z)$ is a decreasing function of $z$.  If $f_+(z)$ vanishes only when $z\to 0$,  then Eq.~(\ref{escale}) implies that $\eta^{-1}$ vanishes for $\rho>\rho_{\rm c}$ only when $\sigma=0$, and so there will be no jamming at finite $\sigma>0$.  If, however, $f_+(z)$ vanishes at some finite $z_0$, then $\eta^{-1}$ will vanish whenever $\sigma/(\rho-\rho_{\rm c})^\Delta = z_0$; there will then be a line of jamming transitions emanating from point {\it J} in the $\rho-\sigma$ plane given by the curve $\rho^*(\sigma)=\rho_{\rm c}+(\sigma/z_0)^{1/\Delta}$.  If $f_+(z)$ vanishes continuously at $z_0$, jamming at finite $\sigma$ will be like a second order transition; if $f_+(z)$ jumps discontinuously to zero at $z_0$, it will be like a first order transition.  Such a first order like transition has been reported in simulations  \cite{Berthier,Varnik} of sheared glasses at finite temperature below the glass transition, $T<T_c$.  However, recent simulations \cite{Ning} of a granular system at $T=0$,  $\rho>\rho_c$, showed that a similar first order like behavior was a finite size effect that vanished in the thermodynamic limit.  With these observations, we leave the question of criticality at finite $\sigma$ to future work

\begin{figure}[tbp]
\epsfxsize=8.6truecm
\epsfbox{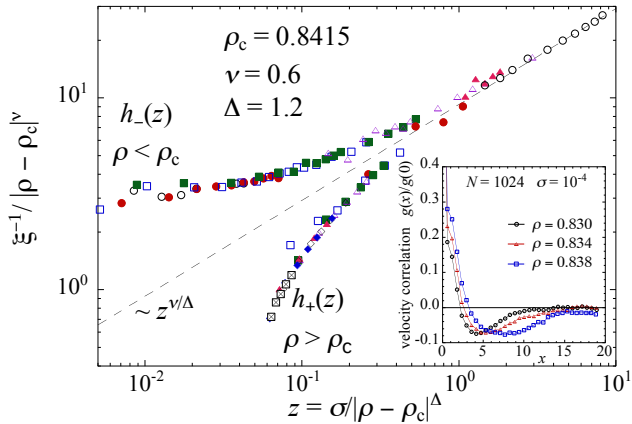}
\caption{(color online) 
Inset: Normalized transverse velocity correlation function $g(x)/g(0)$ vs longitudinal position $x$ for $N=1024$ particles, applied shear stress $\sigma=10^{-4}$, and volume densities $\rho=0.830,$ $0.834$ and $0.838$.  The position of the minimum determines the correlation length $\xi$. 
Main figure: Plot of scaled inverse correlation length $\xi^{-1}/|\rho-\rho_{\rm c}|^\nu$ vs scaled shear stress $z\equiv\sigma/|\rho-\rho_{\rm c}|^\Delta$ for the data of Fig.~\ref{f2}.  We find a good scaling collapse using values $\rho_{\rm c}=0.8415$,  $\Delta=1.2$ (the same as in Fig.~\ref{f3}) and $\nu=0.6$.  Data point symbols correspond to those used in Fig.~\ref{f2}.
}
\label{f4}
\end{figure}

The critical scaling found in Fig.~\ref{f3} strongly suggests that point {\it J} is indeed a true second order phase transition, and thus implies that there ought to be a diverging correlation length $\xi$ at this point.  Measurements of dynamic (time dependent) susceptibilities have been used to argue for a  divergent length scale in both the thermally driven glass transition \cite{Berthier2}, and the density driven jamming transition \cite{Durian2}.
Here we consider the {\it equal time} transverse velocity correlation function in the shear driven steady state, 
\begin{equation}
g(x)=\langle v_y(x_i,y_i)v_y(x_i+x,y_i)\rangle\enspace,
\end{equation}
where $v_y(x_i,y_i)$ is the instantaneous velocity in the $\hat y$ direction, transverse to the direction of the average shear flow, for a particle at position $(x_i,y_i)$.  The average is over particle positions and time.  In the inset to Fig.~\ref{f4} we plot $g(x)/g(0)$ vs $x$ for three different values of $\rho$ at fixed $\sigma=10^{-4}$ and number of particles $N=1024$.  We see that $g(x)$ decreases to {\it negative} values at a well defined minimum, before decaying to zero as $x$ increases.  We define $\xi$ to be the position of this minimum.  That $g(\xi)<0$, indicates that regions separated by a distance $\xi$ are {\it anti-correlated}.  We can thus interpret the sheared flow in the unjammed state as due to the rotation of correlated regions of length $\xi$.  Similar behavior, leading to a similar definition of $\xi$, has previously been found \cite{Tanguy}  in correlations of the nonaffine displacements of particles in a Lennard-Jones glass, in response to small elastic distortions.

As with viscosity, we expect the correlation length $\xi(\rho,\sigma)$ to obey a scaling equation similar to Eq.~(\ref{escale}).  We consider here the inverse correlation length $\xi^{-1}$, which like $\eta^{-1}$ should vanish at the jamming transition, obeying the scaling equation,
\begin{equation}
\xi^{-1}(\rho,\sigma)=|\rho-\rho_{\rm c}|^{\nu}h_\pm\left({\sigma\over|\rho-\rho_{\rm c}|^\Delta}\right)\enspace.
\label{escalexi}
\end{equation}
The correlation length critical exponent is $\nu$, but the crossover exponent $\Delta$ remains the same as for the viscosity.

In Fig.~\ref{f4} we plot the scaled inverse correlation length, $\xi^{-1}/|\rho-\rho_{\rm c}|^\nu$ vs the scaled stress, $\sigma/|\rho-\rho_{\rm c}|^\Delta$.  Using $\rho_{\rm c}=0.8415$ and $\Delta=1.2$, as was found for the scaling of $\eta^{-1}$, we now find a good scaling collapse for $\xi^{-1}$ by taking the value $\nu=0.6$.
By comparing the goodness of the collapse as $\nu$ is varied, we estimate $\nu=0.6 \pm 0.1$.  From the scaling equation Eq.~(\ref{escalexi}) we expect both branches of the scaling function to approach the power law $h_\pm(z)\sim z^{\nu/\Delta}$ as $z\to\infty$, so that $\xi^{-1}\sim \sigma^{\nu/\Delta}$ as $\sigma\to 0$ at $\rho=\rho_{\rm c}$ \cite{nonlinear}.  This is shown as the dashed line in Fig.~\ref{f4}.   Our result is consistent with the conclusion ``$\nu$ is between 0.6 and 0.7" of Drocco et al. \cite{Drocco} for the flowing phase, $\rho<\rho_{\rm c}$.  It also agrees with $\nu=0.71\pm0.08$ found by O'Hern et al. \cite{OHern2} from a finite size scaling argument.  Wyart et al. \cite{Wyart} have proposed a diverging length scale with exponent $\nu=0.5$ by considering the vibrational spectrum of soft modes approaching point {\it J} from the jammed side, $\rho>\rho_c$.  While our results cannot rule out $\nu=0.5$, our scaling collapse in Fig.~\ref{f4} does seem somewhat better when using the larger value $0.6$.

This work was supported by Department of Energy grant DE-FG02-06ER46298 and by the resources of the Swedish High Performance Computing Center North (HPC2N).  We thank J.~P.~Sethna, L.~Berthier, M.~Wyart, J.~M.~Schwarz, N.~Xu, D.~J.~Durian, A.~J.~Liu and S.~R.~Nagel  for helpful discussion.


\begin{thebibliography}{99}

\bibitem{LiuNagel1}{\it Jamming and Rheology}, edited by A.~J.~Liu and S.~R.~Nagel (Taylor \& Francis, New York, 2001).

\bibitem{LiuNagel2}A.~J.~Liu and S.~R.~Nagel, Nature {\bf 396}, 21 (1998).

\bibitem{OHern2}C.~S.~O'Hern et al. Phys. Rev. E {\bf 68}, 011306 (2003).

\bibitem{Durian}D.~J.~Durian, Phys. Rev. Lett. {\bf 75}, 4780 (1995) and Phys. Rev. E {\bf 55}, 1739 (1997).

\bibitem{Makse}H.~A.~Makse, D.~L.~Johnson and L.~M.~Schwartz, Phys. Rev. Lett. {\bf 84}, 4160 (2000).

\bibitem{OHern1}C.~S.~O'Hern et al., Phys. Rev. Lett. {\bf 86}, 000111 (2001) and {\bf 88}, 075507 (2002).

\bibitem{Drocco}J.~A.~Drocco et al., Phys. Rev. Lett. {\bf 95}, 088001 (2005).

\bibitem{Silbert}L.~E.~Silbert, A.~J.~Liu and S.~R.~Nagel, Phys. Rev. Lett. {\bf 95}, 098301 (2005) and Phys. Rev. E {\bf 73}, 041304 (2006).

\bibitem{Ellenbroek}W.~G.~Ellenbroek et al., Phys. Rev. Lett. {\bf 97}, 258001 (2006).

\bibitem{Ning}N.~Xu and C.~S.~O'Hern, Phys. Rev. E {\bf 73}, 061303 (2006).

\bibitem{Schwarz}J.~M.~Schwarz, A.~J.~Liu and L.~Q.~Chayes, Europhys. Lett. {\bf 73}, 560 (2006).

\bibitem{Fisher}C.~Toninelli, G.~Biroli and D.~S.~Fisher, Phys. Rev. Lett. {\bf 96}, 035702 (2006).

\bibitem{Chak}S.~Henkes and B.~Chakraborty, Phys. Rev. Lett. {\bf 95}, 198002 (2005).

\bibitem{Wyart}M.~Wyart, S.~R.~Nagel, T.~A.~Witten, Europhys. Lett. {\bf 72}, 486-492 (2005); M.~Wyart et al., Phys. Rev. E {\bf 72}, 051306 (2005); C.~Brito and M.~Wyart, Europhys. Lett. {\bf 76}, 149 (2006).

\bibitem{Weitz}V.~Trappe et al., Nature (London) {\bf 411}, 772 (2001).

\bibitem{Majmudar}T.~S.~Majmudar et al., Phys. Rev. Lett. {\bf 98}, 058001 (2007).

\bibitem{Durian2}A.~.S.~Keys et al., Nature physics {\bf 3}, 260 (2007).

\bibitem{Swinney}M.~Schr\"{o}ter et al., Europhys Lett. {\bf 78}, 44004 (2007).

\bibitem{Yamamoto}R.~Yamamoto and A.~Onuki, Phys. Rev. E {\bf 58}, 3515 (1998).

\bibitem{Berthier}L.~Berthier and J.-L.~Barat, J.~Chem.~Phys. {\bf 116}, 6228 (2002).

\bibitem{Varnik}F.~Varnik, L.~Bocquet and J.-L.Barrat, J.~Chem.~Phys. {\bf 120}, 2788, (2004).

\bibitem{LeesEdwards}
D.~J.~Evans and G.~P.~Morriss, {\it Statistical Mechanics of Non-equilibrium Liquids} (Academic, London, 1990).

\bibitem{Sasa}T.~Hatano, M.~Otsuki and S.~Sasa, condmat/0607511.

\bibitem{nonlinear} In general, one should consider nonlinear scaling variables.  In our case, the most important correction would be to replace $\rho-\rho_{\rm c}$ in Eq.~(\protect\ref{escale}) by $g_\rho(\rho,\sigma)\equiv \rho-\rho_{\rm c}+c\sigma^2$; this could lead to noticeable corrections to our scaling equation near $\rho=\rho_{\rm c}$.  However, since we find $\Delta>0.5$, our conclusion that $\eta^{-1}\sim\sigma^{\beta/\Delta}$ at $\rho=\rho_{\rm c}$ remains valid.
See, A. Aharony and M.~E.~Fisher, Phys. Rev. B {\bf 27}, 4394 (1983).

\bibitem{Berthier2}L.~Berthier et al., Science {\bf 310}, 1797 (2005).

\bibitem{Tanguy}A.~Tanguy et al., Phys. Rev. B {\bf 66}, 174205 (2002).

\end{thebibliography}
\end{document}